\documentclass[12pt,english]{article}
\pdfoutput=1
\usepackage[T1]{fontenc}
\usepackage{amssymb}
\usepackage{amsmath}
\usepackage{cite}
\usepackage{epsfig}
\usepackage[unicode=true,
 bookmarks=true,bookmarksnumbered=false,bookmarksopen=false,
 breaklinks=false,pdfborder={0 0 1},backref=false,colorlinks=true]
 {hyperref}
\hypersetup{pdfauthor={Grant N. Remmen},
 citecolor=black,linkcolor=black,urlcolor=black}
\usepackage{breakurl}
\usepackage[hang,flushmargin]{footmisc}
\usepackage{breakurl}
\usepackage{color}

\makeatletter
\def\simgt{\mathrel{\lower2.5pt\vbox{\lineskip=0pt\baselineskip=0pt
           \hbox{$>$}\hbox{$\sim$}}}}
\def\simlt{\mathrel{\lower2.5pt\vbox{\lineskip=0pt\baselineskip=0pt
           \hbox{$<$}\hbox{$\sim$}}}}
\makeatother

\newcommand{\be}{\begin{equation}}
\newcommand{\ee}{\end{equation}}
\newcommand{\bea}{\begin{eqnarray}}
\newcommand{\eea}{\end{eqnarray}}

\newcommand{\Fig}[1]{Fig.~\ref{#1}}
\newcommand{\Eq}[1]{Eq.~\eqref{#1}}

\newcommand{\Sec}[1]{Sec.~\ref{#1}}

\newcommand{\App}[1]{App.~\ref{#1}}

\begin{document}

\interfootnotelinepenalty=10000
\baselineskip=18pt
\hfill

\vspace{2cm}
\thispagestyle{empty}
\begin{center}
{\LARGE \bf
Exploration of a Singular Fluid Spacetime
}\\
\bigskip\vspace{1cm}{
{\large Grant N. Remmen}
} \\[7mm]
 {\it Kavli Institute for Theoretical Physics and Department of Physics\\[-1mm]
    University of California, Santa Barbara, CA 93106} \let\thefootnote\relax\footnote{\noindent e-mail: \url{remmen@kitp.ucsb.edu}} \\
 \end{center}
\bigskip
\centerline{\large\bf Abstract}
\begin{quote} \small
We investigate the properties of a special class of singular solutions for a self-gravitating perfect fluid in general relativity: the singular isothermal sphere.
For arbitrary constant equation-of-state parameter $w=p/\rho$, there exist static, spherically-symmetric solutions with density profile $\propto 1/r^2$, with the constant of proportionality fixed to be a special function of $w$. 
Like black holes, singular isothermal spheres possess a fixed mass-to-radius ratio independent of size, but no horizon cloaking the curvature singularity at $r=0$.
For $w=1$, these solutions can be constructed from a homogeneous dilaton background, where the metric spontaneously breaks spatial homogeneity.
We study the perturbative structure of these solutions, finding the radial modes and tidal Love numbers, and also find interesting properties in the geodesic structure of this geometry.
Finally, connections are discussed between these geometries and dark matter profiles, the double copy, and holographic entropy, as well as how the swampland distance conjecture can obscure the naked singularity.
\end{quote}

\setcounter{footnote}{0}

\newpage
\tableofcontents
\newpage

\section{Introduction}
\vspace{-1.5mm}

The construction and detailed study of solutions to the Einstein equations has played an important role in a century of progress in physics.
Work on discovering and characterizing spacetime geometries is of particular relevance in many areas of active research ranging from holography and quantum gravity, to scattering amplitudes and the double copy, to metrics for compact astrophysical objects.

In this work, we will investigate a set of solutions for a self-gravitating perfect fluid in general relativity: the singular isothermal sphere (SIS). 
The solution itself dates back to Refs.~\cite{Tolman:1939jz,Oppenheimer:1939ne}, and it is important in a variety of astrophysical applications~\cite{MTW,Weinberg,SmollerTemple,ShuNewtonian,CaiShu2005,OriPiran}.
The SIS possesses many interesting properties, including a naked singularity in curvature, fixed mass-to-radius ratio like black holes, and the ability to trap light.
Examples of these solutions can occur in string theory and are relevant for the gravity/Yang-Mills double copy.
As we will see, the singularity present in these solutions exhibits potential connections to the swampland distance conjecture.
Though the SIS solutions have been previously considered in astrophysical contexts, various features remain to be studied, which we will explore in this work.

This paper is organized as follows. In \Sec{sec:solution}, we will review the Tolman-Oppenheimer-Volkhoff (TOV) equations describing a relativistic fluid, as well as review the construction of the SIS, calculate the stress tensor and surface gravity of the
boundary shell, and give an example of these solutions that occurs Einstein-dilaton gravity.
We treat perturbations in \Sec{sec:perturbations}, including both time-dependent radial modes and static, quadrupole tidal force solutions that allow for computation of the electric Love number.
We consider connections and applications in \Sec{sec:discussion}---including the rotation curve and relevance for dark matter, outer entropy, the double copy, and implications of the swampland distance conjecture for cosmic censorship---and in \Sec{sec:conclusions} discuss future directions.

In the appendices, other interesting properties of SIS spacetimes are investigated.
In \App{sec:singularity}, we consider the singularity at $r=0$, showing that the curvature diverges, that it is uncloaked by a horizon, and that it can be reached in finite proper time.
The existence of such a singularity is to be expected given the boundary conditions of our solution~\cite{Anastopoulos:2020mrt}; see also Refs.~\cite{Christodoulou,Joshi:1993zg}.
In \App{sec:geodesics}, we study orbits in these geometries, using the geodesic equation to construct an effective potential in a new, effective radial coordinate.
While the dynamics of fluid sphere solutions has been well investigated in the context of collapse~\cite{ShuNewtonian,CaiShu2005,Herrera:1980zz,Herrera1990,Aguirre,OriPiran}, a general treatment of the behavior of geodesics within the static, singular SIS geometry has been less studied.
Orbits in these geometries exhibit remarkable features, including equation-of-state-dependent precession, photon spheres, and trapped spiraling null geodesics.

\section{Construction}\label{sec:solution}

\subsection{TOV and SIS}

Let us first briefly review the equations of hydrostatic equilibrium for a general relativistic, self-gravitating perfect fluid.
We start with a metric ansatz of the form
\be
{\rm d}s^2 =  -e^{2\Phi(r)} {\rm d}t^2 + \left[1-\frac{2m(r)}{r}\right]^{-1}{\rm d}r^2 + r^2 {\rm d}\Omega^2,
\ee
describing a static, spherically-symmetric spacetime (i.e., of Petrov type D). 
To satisfy the Einstein equations\footnote{We set the gravitational constant $G = M_{\rm Pl}^{-2}$ to unity throughout, since it can be straightforwardly restored by dimensional analysis.} for a static perfect fluid,
\be
\frac{1}{8\pi}\left(R_{\mu\nu} - \frac{1}{2}Rg_{\mu\nu}\right) = T_{\mu\nu} = (\rho + p)u_\mu u_\nu +p g_{\mu\nu},
\ee
where $u$ is a unit timelike vector, we must satisfy the TOV equations~\cite{Tolman:1939jz,Oppenheimer:1939ne}:
\be 
\begin{aligned}
\frac{{\rm d}p}{{\rm d}r}&=-(\rho+p)\frac{{\rm d}\Phi}{{\rm d}r}\\
\frac{{\rm d}m}{{\rm d}r}&=4\pi r^2 \rho \\
\frac{{\rm d}\Phi}{{\rm d}r} &=\frac{m + 4\pi r^3 p}{r(r-2m)},
\end{aligned}
\ee
The function $m(r)$ is the mass\footnote{Defining a quasilocal ``mass'' enclosed by a surface is a notoriously subtle problem in general relativity~\cite{Szabados:2004vb}; in this case, $m(r)$ corresponds to the mass quantities of Hawking~\cite{Hawking:1968qt,Hayward:1993ph}, Misner, Sharp, and Hernandez~\cite{Misner:1964je,Hernandez:1966zia}, and Cahill and McVittie~\cite{CahillMcVittie}.} (in Planck units) within a radius $r$. The TOV equations are a system of nonlinear, coupled ordinary differential equations with a rich solution structure, which has been explored for decades in the modeling of relativistic stars.

We will construct a rather unique set of solutions of the TOV equations that will exhibit many remarkable properties: the SIS.
This geometry is a special case of the Tolman VI solution~\cite{Tolman:1939jz} and has proven useful in various gravitational and astrophsical contexts; the SIS solution appears for the particular case of an ultrarelativistic ($p=\rho/3$) fluid in Refs.~\cite{Oppenheimer:1939ne,MTW,Weinberg} and was also considered in the context of Newtonian collapse in Ref.~\cite{ShuNewtonian}, relativistic collapse in Ref.~\cite{CaiShu2005,OriPiran}, and as a portion of a shock wave geometry in Ref.~\cite{SmollerTemple}.
However, there are various aspects of this solution, including its perturbation theory, that have not as yet been fully explored and which will prove interesting.

Let us choose as our ansatz the requirement of a fixed equation of state, $p = w\rho$ for some constant parameter $w$.  We further impose the density profile $\rho = C/4\pi r^2$, where $C$ is some as-yet-undermined constant. Enforcing the weak energy condition, we require $w\geq -1$ and $C \geq 0$ and fix the integration constant in $m$ so that $m(0)$ vanishes. Remarkably, the TOV equations can then be solved for a single value of the density constant,
\be
C = \frac{2w}{1+6w+w^2},\label{eq:C}
\ee
giving us the solution:
\be
\begin{aligned}
\rho = p/w &= \frac{w}{2\pi r^2 (1+6w+w^2)}\\
m(r) &= \frac{2wr}{1+6w+w^2}\\
\Phi &= \frac{2w}{1+w}\log[(1+w)r] + \Phi_0.
\end{aligned}\label{eq:solution}
\ee
For arbitrary $w$, no single power-law $\rho \propto r^n$ yields a solution except for $n=2$.
The solution in \Eq{eq:solution} is striking in that its density profile is absolutely fixed for given $w$.
This is in stark contrast with typical solutions of the TOV equations, where one fixes the central density arbitrarily as a boundary condition~\cite{MTW}.
From \Eq{eq:solution}, we see that requiring nonnegative $\rho$ implies that we must consider $w \geq 0$.

\subsection{Connecting to Schwarzschild}\label{sec:junction}

Let us join this solution with the Schwarzschild vacuum solution for $r>R$, where $R$ is the radius of our object.
We define the total mass as
\be
M = m(R) =  \frac{2wR}{1+6w+w^2}.\label{eq:M}
\ee
We can set the integration constant $\Phi_0$ (equivalently, define the scale of our internal $t$ coordinate) such that $g_{tt}(R) = 1-2M/R$.
Putting everything together, we have the metric:
\be
 {\rm d}s^{2}  =
 \begin{cases}
 \displaystyle
-\frac{(1+w)^{2}}{1+6w+w^{2}}\left(\frac{r}{R}\right)^{\frac{4w}{1+w}}{\rm d}t^{2}+\frac{1+6w+w^{2}}{(1+w)^{2}}{\rm d}r^{2}+r^{2}{\rm d}\Omega^{2} & r\leq R\\
\displaystyle
-\left(1-\frac{2M}{r}\right){\rm d}t^{2}+\left(1-\frac{2M}{r}\right)^{-1}{\rm d}r^{2}+r^{2}{\rm d}\Omega^{2} & r>R.
\end{cases}\label{eq:metric}
\ee
The geometry in \Eq{eq:metric} solves the Einstein equations describing our fluid or vacuum for $r<R$ and $r>R$ respectively, but not at the surface $\Sigma$ located at $r=R$.
Since the metric is continuous but not smooth at $\Sigma$, the junction conditions will require supplementing the energy-momentum tensor with an additional shell with support on $\Sigma$, as we will discuss.
Such a sharp cutoff was also considered in the collapsing case in Refs.~\cite{CaiShu2005,OriPiran}, as well as a smoothed generalization in Ref.~\cite{OriPiran}.

\begin{figure}[t]
\begin{center}
\hspace{-15mm}\includegraphics[width=10.5cm]{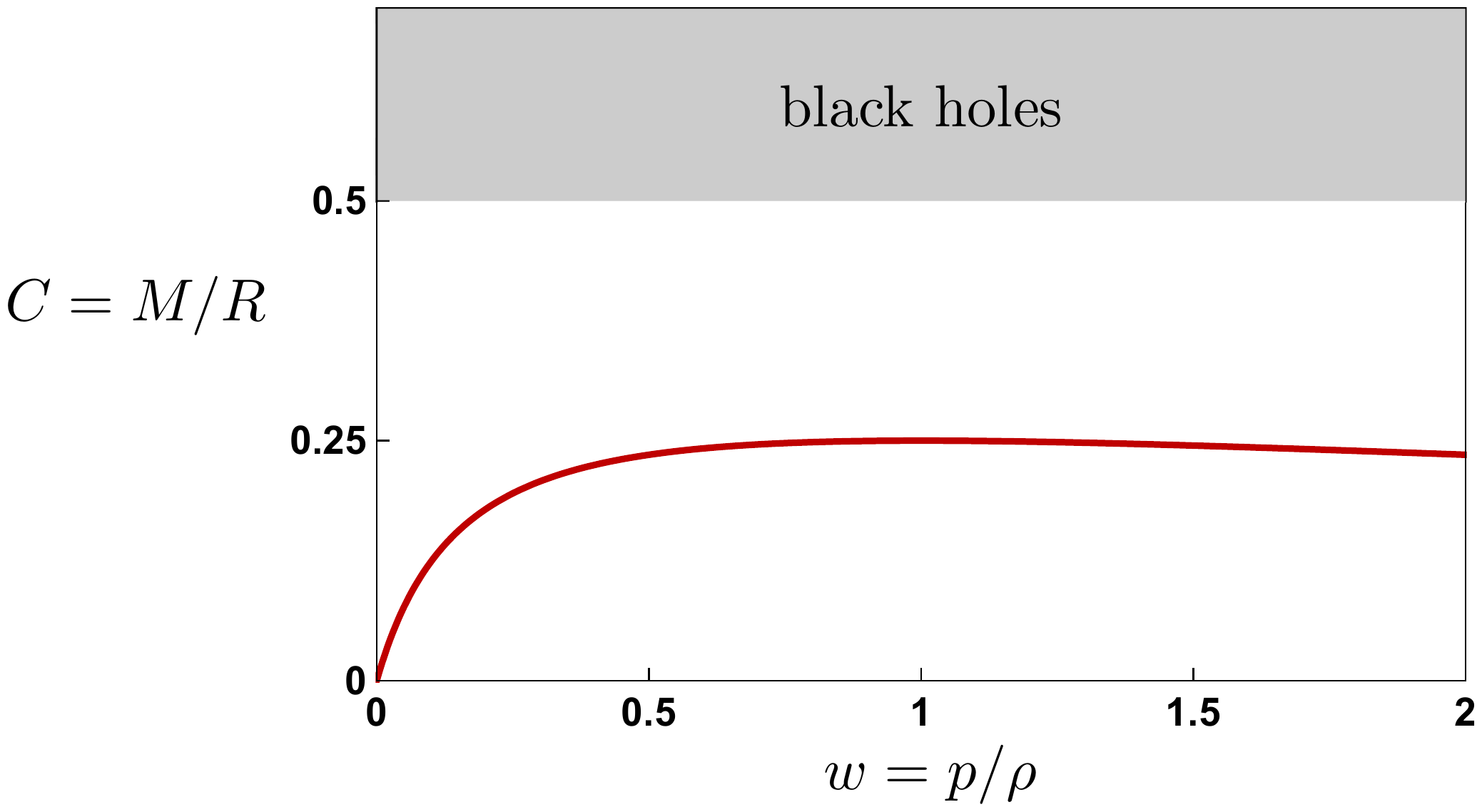}
\end{center}\vspace{-7mm}
\caption{Compactness parameter $C$ for the singular fluid solutions in \Eq{eq:metric}.
}
\label{fig:compactness}
\end{figure}

We first note from \Eq{eq:M} that $C$ as defined in \Eq{eq:C} is the compactness parameter, $C=M/R$,
so that $C\geq 1/2$ would correspond to the formation of a black hole; we find that for this class of solutions, we have $C \leq 1/4$, with the maximum achieved for stiff matter with $w=1$ (see \Fig{fig:compactness}).
Strikingly, $C$ is invariant under the transformation $w\rightarrow 1/w$.
Since requiring a causal speed of sound implies $w\leq 1$, we will restrict to $w\in [0,1]$ henceforth.
Due to the form of the solution in \Eq{eq:solution}, the compactness parameter for the matter interior to any given radius $r\leq R$ is a constant.
This fluid solution therefore has no horizons~\cite{CaiShu2005}.
Despite its horizonless nature, however, this geometry bears some striking resemblances to a black hole. 
For given $w$, the ratio $M/R$ is a fixed, order-one number, just like black holes (and comparable to the $C$ values one expects for neutron stars).
Moreover, as we will see in \App{sec:singularity}, like black holes the fluid solution exhibits a curvature singularity at $r=0$.

At $\Sigma$, where the fluid interior joins the Schwarzschild vacuum exterior, there must be a thin shell of matter that confines the fluid.
The reason for this is as follows.
Taking the piecewise-defined metric in \Eq{eq:metric} and evaluating its Ricci tensor in the limit $r\rightarrow R$, the discontinuities in metric derivatives give rise to a contribution to $R_{\mu\nu}$ with delta-function support at $r=R$.
Plugging this into the Einstein equations gives us the needed extra shell of matter on $\Sigma$ implied by the Israel junction conditions.
The induced metric on $\Sigma$ is given by $\gamma_{ab}{\rm d}x^a {\rm d}x^b = - (1-2C){\rm d}t^2 + R^2 {\rm d}\Omega^2$.
By explicit calculation, one finds that the matter density within the shell can be written as $T_{ab} = -\varsigma (u_a u_b + \gamma_{ab} )$,
corresponding to a perfect fluid with surface tension (negative pressure) $\varsigma = C w/4\pi R = R \,p(R)$.
In the full spacetime coordinates, the extra energy-momentum tensor of the shell is 
\be 
\Delta T_{\mu\nu} = -\varsigma\,\delta(r-R)\,\times {\rm diag}(0,0,R^2,R^2\sin^2\theta).\label{eq:surface}
\ee
This is similar to the thin shells providing the required tension in the gravastar~\cite{Mazur:2001fv}.
While the energy-momentum tensor of the surface in \Eq{eq:surface} violates the null energy condition, this can be simply rectified by adding an arbitrary surface density larger than $\varsigma$. The only effect on the geometry of doing so would be to shift $M$ in the $r>R$ part of the metric~\eqref{eq:metric} by the total mass of the shell.

The metric in \Eq{eq:metric} possesses a timelike Killing vector $t^\mu$, $\nabla_{(\mu}t_{\nu)} = 0$, proportional to $u^\mu$, with normalization such that $t^\mu \partial_\mu = \partial/\partial t$.
The orbits of this Killing vector, hypersurface-orthogonal to the constant-$t$ surface, describe the spacetime trajectory of an observer at fixed spatial position $(r,\theta,\phi)$.
The acceleration of such a trajectory is $a^\mu = (t^\nu \nabla_\nu t^\mu)/|t|^2$, where 
\be
|t| \equiv \sqrt{-t_\mu t^\mu} = 
\begin{cases}
(1+w)\sqrt{\frac{M}{2Rw}} \left(\frac{r}{R}\right)^{\frac{2w}{1+w}} & r<R \\
\sqrt{1-\frac{2M}{r}} & r \geq R
\end{cases}
\ee
is the redshift factor. Writing $a=\sqrt{a_\mu a^\mu}$, we have
\be 
a =  
\begin{cases}
\sqrt{\frac{2wM}{R}}\frac{1}{r} & r<R \\
\frac{M}{r^2} \left(1-\frac{2M}{r}\right)^{-1/2} & r \geq R.
\end{cases}
\ee
The surface gravity of the object is the force at infinity needed to suspend an observer at the $r=R$ surface. Taking this limit from the inside versus outside yields a discontinuity:
\be
\begin{aligned}
\kappa_+ &= \lim_{r\rightarrow R^+} |t| a = \frac{M}{R^2}\\
\kappa_- &= \lim_{r\rightarrow R^-} |t|a = (1+w)\kappa_+.
\end{aligned} 
\ee
The difference across the $r=R$ boundary,
\be
\Delta \kappa = \kappa_+ - \kappa_- = -w\kappa_+ = -\frac{Cw}{R} = - 4\pi \varsigma, 
\ee
accounts for the extra force needed to counteract the tension supplied by the boundary shell.
A similar effect was found in Ref.~\cite{Mazur:2015kia} in the context of a constant-density fluid model of a compact object.
Letting the size of the SIS become arbitrarily large, by sending $R\rightarrow \infty$, would lead to a divergent mass by \Eq{eq:M} and hence would be incompatible with asymptotically flat spacetime, as noted in Refs.~\cite{ShuNewtonian,CaiShu2005,OriPiran}.

\subsection{Einstein-dilaton gravity}

We can realize the $1/r^2$ fluid solution given by \Eq{eq:metric} within concrete field theory examples.  Taking the theory of Einstein gravity plus a massless, minimally coupled scalar,
\be
{\cal L} = \frac{1}{16\pi} R - \frac{1}{2}(\partial\Phi)^2, \label{eq:Einsteinscalar}
\ee
with equations of motion,
\be 
\begin{aligned}
R_{\mu\nu} - \frac{1}{2}Rg_{\mu\nu} &= 8\pi \left[\partial_\mu \Phi \partial_\nu \Phi - \frac{1}{2}g_{\mu\nu}(\partial\Phi)^2 \right]\\
\Box \Phi &= 0,
\end{aligned}\label{eq:Einsteinscalareoms}
\ee
we find that the metric in \Eq{eq:metric} is a solution, for the choice $w=1$,
\be 
{\rm d}s^2 = r^2 \left(-\frac{{\rm d}t^2}{2R^2}+{\rm d}\Omega^2\right)+2\,{\rm d}r^{2}, \label{eq:metscalar}
\ee
with the scalar satisfying a spatially homogeneous profile,
\be 
\Phi = t/4\sqrt{\pi}R.\label{eq:phisol}
\ee
Remarkably, despite the spatial homogeneity of the scalar background, the metric in \Eq{eq:metscalar} spontaneously breaks this symmetry, becoming singular at the origin.
The gravity-plus-dilaton theory in \Eq{eq:Einsteinscalar} appears in string theory in the Einstein frame; the metric in \Eq{eq:metscalar} is thus relevant to study as a string theory solution.

\section{Perturbations}\label{sec:perturbations}

To further understand the SIS solutions, we now would like to investigate how they behave under radial and tidal deformations.
We consider a perturbation of the metric, sending $g_{\mu\nu} \rightarrow g_{\mu\nu} + h_{\mu\nu}$, with the Einstein tensor transforming as $G_\mu^{\;\;\nu}\rightarrow G_\mu^{\;\;\nu} + \delta G_\mu^{\;\;\nu}$, where at linear order,
\be 
\begin{aligned}
\delta G_\mu^{\;\;\nu} &=-R_{\mu\alpha}h^{\alpha\nu} + \frac{1}{2}\delta_\mu^\nu R_{\alpha\beta}h^{\alpha\beta} \\
&\qquad +  \frac{1}{2}\left(\nabla_\alpha \nabla_\mu h^{\nu\alpha}  +  \nabla^\alpha\nabla^\nu h_{\mu\alpha}  -  \nabla^\nu \nabla_\mu h_\alpha^{\;\;\alpha} - \delta_\mu^\nu \nabla_\alpha\nabla_\beta h^{\alpha\beta} -\Box h_\mu^{\;\;\nu}  +  \delta_\mu^{\nu}\Box h_\alpha^{\;\;\alpha}\right) .
\end{aligned}
\ee
While in this paper we will restrict to consideration of gravitational effects, in astrophysical applications various thermal effects---radiation, streaming, etc.---can play an important role in considerations of stablity; see Refs.~\cite{Herrera:1980zz,Herrera1990,Aguirre}.

\subsection{Radial perturbations}
Let us first consider radial perturbations, taking as our ansatz
\be
h_{\mu\nu} = {\rm diag}(-g_{tt}(r) h_0(t,r) ,g_{rr}(r) h_2(t,r),0,0)
\ee
for the metric perturbation, with $g_{tt}$ and $g_{rr}$ given by the unperturbed $r<R$ solution in \Eq{eq:metric}. We allow the pressure, density, and velocity field to be perturbed by $\delta p(t,r)$, $\delta \rho(t,r)$, and $\delta u_\mu(t,r)$, respectively, so that 
\be 
\delta G_\mu^{\;\;\nu} = 8\pi \delta T_\mu^{\;\;\nu},\label{eq:einpert}
\ee
where
\be 
 \delta T_\mu^{\;\;\nu} = (\delta \rho + \delta p) u_\mu u^\nu + \delta p\,\delta_\mu^\nu + (\rho+p)(u_\mu \delta u^\nu + \delta u_\mu \,u^\nu),
 \ee
where $\delta u^\mu$ is defined by contracting $\delta u_\mu$ with the inverse of the background metric. We will require $\delta p = w\,\delta \rho$, so that the equation of state remains fixed, and further $u_\mu \delta u^\mu = 0$ so that $u^2 = -1$ is unchanged.
Preserving spherical symmetry, we need consider only $\delta u_r$ in the velocity perturbation.

\begin{figure}[t]
\begin{center}
\includegraphics[width=11cm]{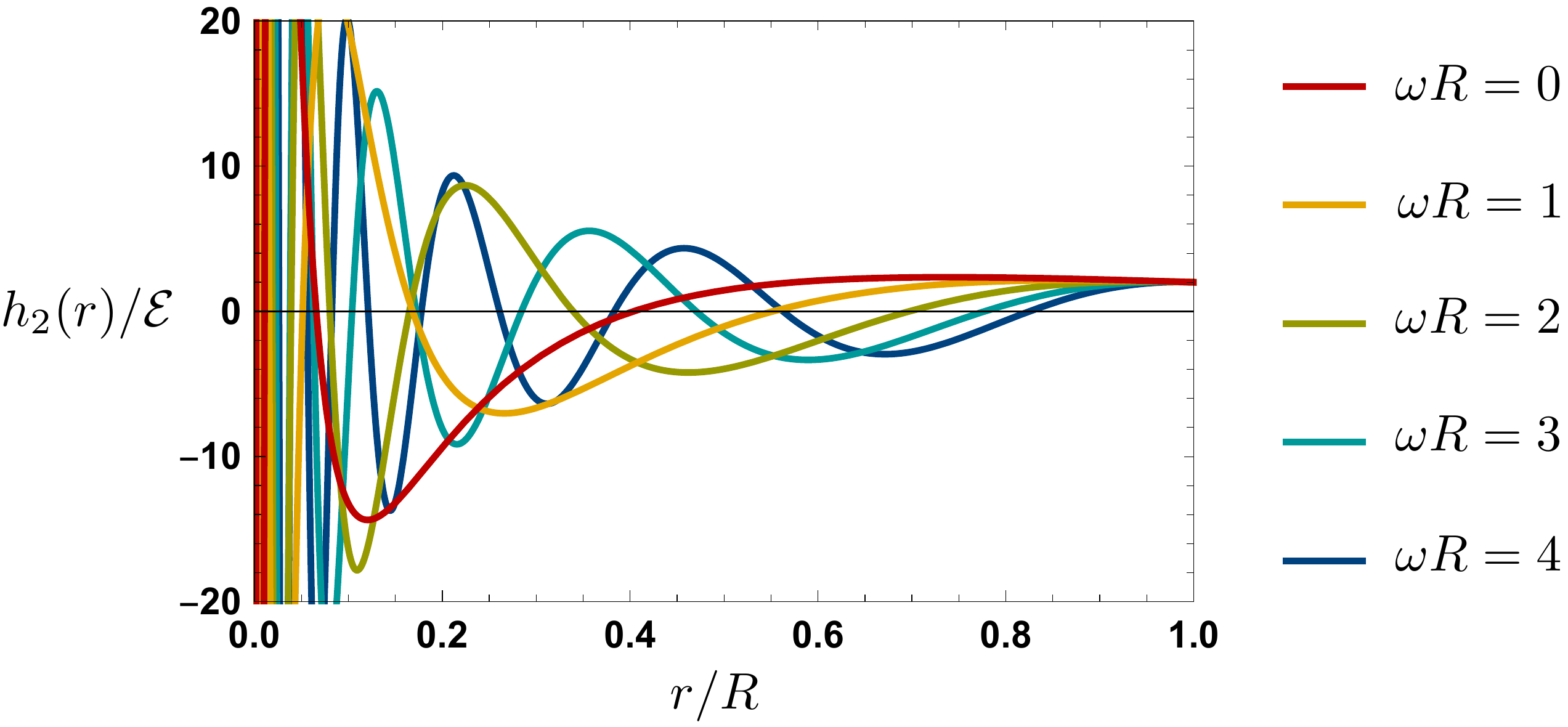}
\end{center}\vspace{-7mm}
\caption{Solutions for the radial perturbation given in \Eq{eq:sols} in the stiff fluid ($w=1$) case, for various choices of frequency $\omega$, exhibiting infinitely many oscillations as the singularity at $r=0$ is approached.
}
\label{fig:perturbation}
\end{figure}

Considering perturbations that are eigenfunctions with respect to the Killing vector $\partial_t$, we factor out the time dependence as 
\be
\begin{aligned}
h_0(t,r) &= e^{i\omega t} h_0(r) \\
h_2 (t,r) &= e^{i\omega t} h_2(r)\\
\delta p(t,r) &= e^{i\omega t} \delta p(r)\\
\delta u_r(t,r) &= e^{i\omega t} \delta u_r(r).
\end{aligned} 
\ee
Requiring $\delta G_{t}^{\;\;t} = 8\pi \delta T_{t}^{\;\;t}$ and $\delta G_{t}^{\;\;r} = 8\pi \delta T_{t}^{\;\;r}$ gives us, respectively, $\delta p$ and $\delta u_r$, written in terms of $h_2(r)$ and its derivative.
Similarly, enforcing the $rr$ component of \Eq{eq:einpert} yields $h_0'(r)$ in terms of $h_2(r)$ and its derivative, so that we find:
\be
\begin{aligned}
h_0'(r) &= -\frac{2w}{(1+w)Cr} h_2(r) - w  h_2'(r)\\
\delta p(r) &=\frac{C}{16\pi r^2} (1+w)^2 [ h_2(r) + r  h'_2(r)]\\
\delta u_r(r) &= i\omega r \left(\frac{r}{R}\right)^{-\frac{2w}{1+w}}\sqrt{\frac{w}{2C^3}}\frac{h_2(r)}{(1+w)^2}.
\end{aligned} 
\ee
Finally, the last remaining nontrivial components of \Eq{eq:einpert} ($\theta\theta$ and $\phi\phi$) imply an ordinary differential equation for $\delta h_2(r)$:
\be 
\frac{1}{2} C (1+w)^4 r^2 h_2''(r) + C (1+w)^3 (1+2w) r h_2'(r) + 2w\left[(1+w)^2 + \frac{\omega^2 R^2}{C}\left(\frac{r}{R}\right)^{\frac{2(1-w)}{1+w}} \right]h_2(r) = 0.
\ee
This differential equation is of Bessel type and has general solution in terms of Bessel functions of pure imaginary order, $J_{i\alpha}(x)$:
\be
\begin{aligned}
h_2(r) &= \left(\frac{r}{R}\right)^{-\frac{1+3w}{2(1+w)}}\left\{c_+ J_{+i\frac{\sqrt{7 +42 w - w^2}}{2 (1 - w)}}\left[\tfrac{\omega R (1+6w+w^2)}{\sqrt{w}(1-w^2)} \left(\tfrac{r}{R}\right)^{\frac{1-w}{1+w}}\right] \right. \\
&\qquad\qquad\;\;\;\;\;\; \left. + c_-J_{-i\frac{\sqrt{7 +42 w - w^2}}{2 (1 - w)}}\left[\tfrac{\omega R (1+6w+w^2)}{\sqrt{w}(1-w^2)} \left(\tfrac{r}{R}\right)^{\frac{1-w}{1+w}}\right]\right\}.\label{eq:h2sol}
\end{aligned}
\ee
In the $w=1$ case, the solution is
\be
h_2(r) =  c_+ \left(\frac{r}{R}\right)^{-1+i\sqrt{3+4R^2\omega^2}} + c_- \left(\frac{r}{R}\right)^{-1-i\sqrt{3+4R^2\omega^2}} .\label{eq:h2solw1}
\ee
Now, to enforce Neumann boundary conditions with $\delta p = 0$ at the surface of the star, we set $|c_+| = |c_-|$, with complex phase chosen to cancel the relative phase between the complex Bessel functions. For example, setting $c_+ = c_- = {\cal E}$ in the $w=1$ case given in \Eq{eq:h2solw1}, we have
\be
\begin{aligned}
h_0(r) &= -\frac{2R{\cal E}}{(1+\omega^2 R^2)r}\left\{\omega^2 R^2 \cos\left[\sqrt{3+4\omega^2 R^2} \log(r/R) \right]\right. \\& \qquad\qquad\qquad\qquad \left. + \sqrt{3+4\omega^2 R^2}\sin\left[\sqrt{3+4\omega^2 R^2} \log(r/R) \right]\right\}\\
h_2(r) &= \frac{\sqrt{2}}{i \omega R} \delta u_r(r) =  \frac{2R{\cal E}}{r} \cos\left[\sqrt{3+4\omega^2 R^2} \log(r/R) \right]\\
\delta p(r) &= - \frac{R{\cal E}}{8\pi r^3}\sqrt{3+4\omega^2 R^2}\sin\left[\sqrt{3+4\omega^2 R^2} \log(r/R) \right].
\end{aligned}\label{eq:sols}
\ee
See \Fig{fig:perturbation} for an illustration.
For a typical fluid solution, e.g., a neutron star, one would fix additional boundary conditions at the origin, resulting in quantized values of $\omega$: the normal modes.
For our fluid solution, however, the perturbations have an essential singularity at $r=0$, as one can see from \Eq{eq:sols} and \Fig{fig:perturbation}, preventing the fixing of boundary conditions there; as a result, $\omega$ is undetermined.
This is a manifestation of the breakdown in predictability generated by the singularity discussed in \App{sec:singularity}.

\subsection{Love numbers}
\begin{figure}[t]
\begin{center}
\includegraphics[width=9cm]{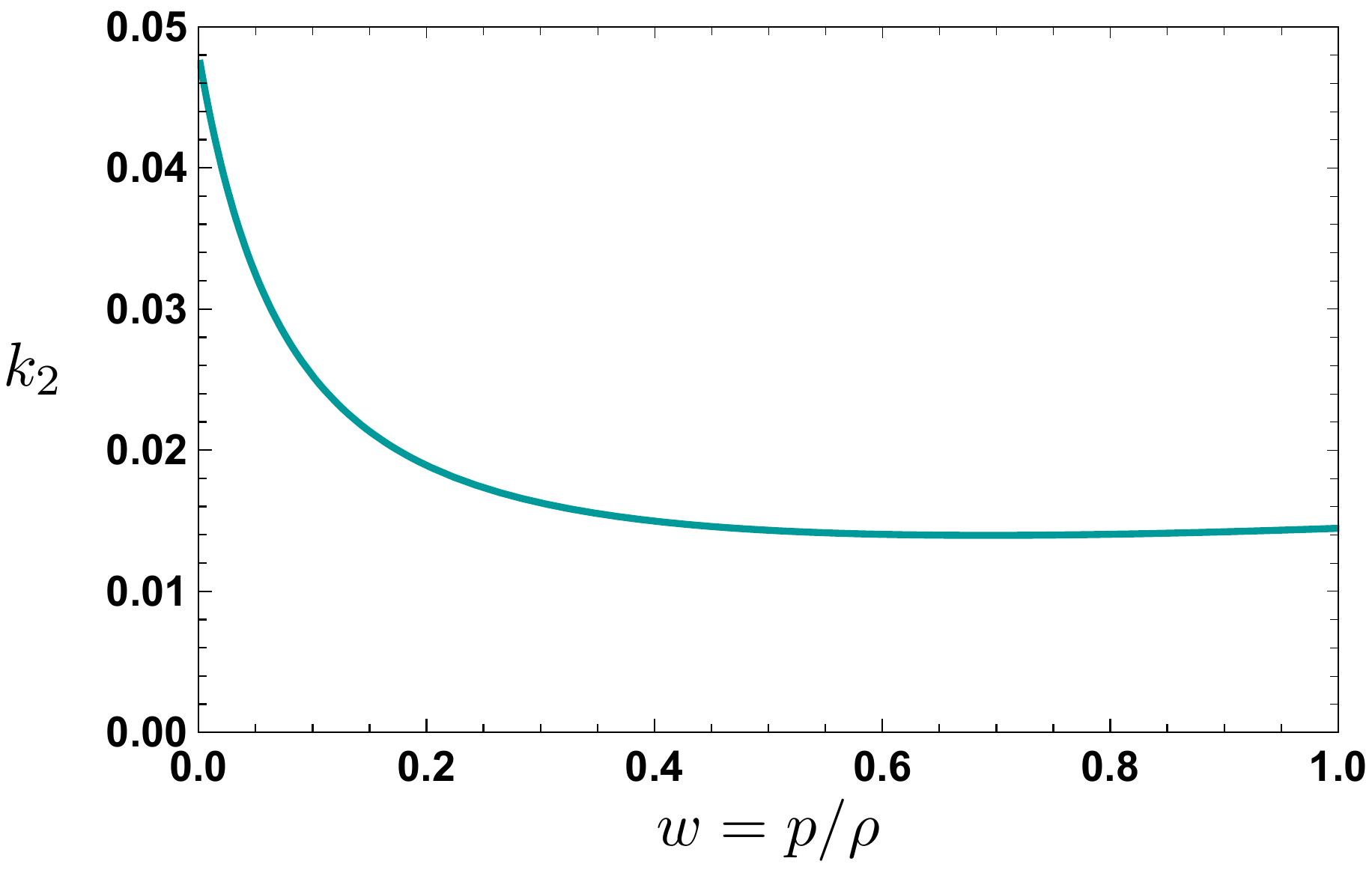}
\end{center}\vspace{-7mm}
\caption{Quadrupole electric Love number $k_2$ for the SIS, computed in \Eq{eq:k2} as a function of equation-of-state parameter $w$.
}
\label{fig:LoveNumber}
\end{figure}

We now consider a nonspherical perturbation of the metric. Specifically, let us consider a static, parity-even deformation, through which we will be able to model the response of our solution to an external tidal field. In Regge-Wheeler gauge~\cite{Regge:1957td,Thorne}, we write
\be 
h_{\mu\nu} =Y_{\ell 0}(\theta,\phi) \times {\rm diag} (-g_{tt}(r) H_0(r), g_{rr}(r) H_2 (r), K(r), K(r) \sin^2\theta ).
\ee
Taking the energy-momentum tensor to be $\delta T_\mu^{\;\;\nu} = (\delta \rho  + \delta p) u_\mu u^\nu + \delta p \delta_\mu^\nu$, where $\delta p = w\,\delta \rho$, 
we find that demanding $\delta G_\mu^{\;\;\nu} = 8\pi \delta T_\mu^{\;\;\nu}$ allows us to eliminate $\delta p$ as in Ref.~\cite{Hinderer:2007mb}. For $r<R$, we find the solution:
\be
H(r) \equiv H_0(r) = H_2(r) = c_1 r^{n_+} + c_2 r^{n_-},
\ee
where
\be
n_\pm = \frac{-1-3w \pm \sqrt{w^2-42w-7 + \frac{8\ell(\ell+1)}{C}w}}{2(1+w)}
\ee
and $K'$ is fixed in terms of $H(r)$ by the equations of motion. Imposing regularity of $h_{\mu\nu}$ as $r\rightarrow 0$ requires $c_2 = 0$.
Meanwhile, for $r>R$, we have
\be
H(r) \equiv H_0(r) = H_2(r) = c_3 P_\ell^2\left(\frac{r}{M}-1\right)+c_4 Q_\ell^2\left(\frac{r}{M}-1\right),
\ee
where $P_\ell^n$ and $Q_\ell^n$ are the associated Legendre functions of the first and second kind, respectively. We can fix $c_3$ and $c_4$ as in Ref.~\cite{Hinderer:2007mb} by requiring continuity of $H(r)$ and $H'(r)$ at $r=R$.
Rescaling $c_1$ to canonical normalization, we have the expansion of $g_{tt}$ at large $r$ for the quadrupole ($\ell=2$) case:
\be
g_{tt} = -1+\frac{2M}{r}  + {\cal E} \left[(r-2M)^2 + \frac{2R^5 k_2}{r^3}\right] + O((M/r)^4).
\ee
Here, we have found the (electric) tidal Love number $k_2$ to be:
\be
k_2 =  \frac{\displaystyle 8C^5}{5\left\{\displaystyle \frac{C^2[(1+w)^2 (1+4w+w^2) p_1(w)n_+-2p_2(w)]}{ w(1+w)^4[(1+w)^2 n_+ -2(1+4w+w^2)]} + 3 \log(1-2C)\right\}},\label{eq:k2}
\ee
where we have defined
\be
\begin{aligned}
p_1 (w) &= 3 + 24w + 34w^2 + 24w^3 + 3w^4\\
p_2 (w)&=3+48w+280w^2 + 752w^3 + 1066w^4+752w^5+280w^6+48w^7+3w^8.
\end{aligned} 
\ee
Positivity of $k_2$, depicted in \Fig{fig:LoveNumber}, is consistent with stability of these objects; negative $k_2$ would have indicated a nonspherical instability~\cite{1110.3764}.  Higher-order Love numbers can be computed similarly.
At small $w$, which corresponds to the Newtonian (small-$C$, weak-field) limit, we have
\be 
k_2 = \frac{1}{8}(21-5\sqrt{17})(1-10w) + O(w^2),
\ee
while for $w=1$, we have $k_2 =  1/[40(35-48\log 2)]\approx 0.01446$.

\section{Discussion}\label{sec:discussion}
The metrics described by \Eq{eq:metric} have various other interesting properties that we consider, including the rotation curve and applications to galactic dark matter, holographic outer entropy,  connections to the double copy between gravity and Yang-Mills theory, and implications for the swampland distance conjecture.

\subsection{Rotation curve}\label{sec:rotationcurve}

The $1/r^2$ density profile in \Eq{eq:solution} is reminiscent of the ansatz for a galactic dark matter distribution that produces a flat rotation curve in Newtonian gravity.
Of course, in order to describe a realistic dark matter distribution, one would need to implement the transition from the fluid region to Schwarzschild in a smoother way than in \Sec{sec:junction}, but for the present work, we will illustrate the salient features of the interior, fluid region.
For a circular orbit ($\theta = \pi/2$, $\dot r = 0$) at $r<R$ in the metric in \Eq{eq:metric}, the geodesic equation~\eqref{eq:geos} has solution 
\be
\begin{aligned}
t &=  \frac{\tau}{\sqrt{1-2C}}\left(\frac{r}{R}\right)^{-\frac{2w}{1+w}} \\
\phi &= \sqrt{\frac{2w}{1+w}}\frac{\tau}{r}.
\end{aligned} 
\ee
The proper orbital velocity is thus independent of $r$, just as in the Newtonian case:
\be
v = r \dot\phi = \sqrt{\frac{2w}{1+w}},\label{eq:rotationalvelocity}
\ee
while the apparent orbital velocity $v/\dot t$, as measured in the asymptotic time coordinate $t$, vanishes as $r\rightarrow 0$.

In the small-$w$ limit, we can apply the results of the nonrelativistic kinetic theory of gases to check consistency of \Eq{eq:rotationalvelocity}.
Virializing the system, the equipartition theorem implies that we can relate the rotational velocity to the average kinetic energy per particle, $3\sigma^2/2=v^2/2$, where $\sigma^2$ is the velocity dispersion along any direction; this can be written as a temperature $T=3 m_0 \sigma^2/2$, where $m_0$ is the particle mass.
Meanwhile, the nonrelativistic ideal gas law can be written as $w = p/\rho =T/m_0$, so we should have $v^2/2 = w$.
And indeed, expanding \Eq{eq:rotationalvelocity} for small $w$, this is precisely what we find.

\subsection{Outer entropy}

The metric in \Eq{eq:metric} presents an opportunity to apply the outer entropy formalism of Refs.~\cite{Nomura:2018aus,Bousso:2018fou}.
Given a codimension-two surface $\chi$, the outer entropy~\cite{Engelhardt:2017aux} is the area (times $1/4G\hbar$) of the largest HRT surface that can be constructed in the spacetime subject to keeping the geometry in the outer causal wedge of $\chi$ fixed, along with the dominant energy condition.
More heuristically, given a normal (i.e., non-trapped) two-dimensional, closed surface in four spacetime dimensions, the outer entropy is the area of the largest black hole that can be fit inside the surface while holding the geometry outside the surface fixed.
This quantity is a useful, manifestly nonnegative, holographic coarse-grained information-theoretic measure and has been shown to have beautiful area law properties~\cite{Engelhardt:2017aux,Nomura:2018aus,Bousso:2018fou}.

Let us take $\chi$ at fixed $r<R$ and define the ingoing and outgoing orthogonal null congruences,
\be
\begin{aligned}
k_\pm^\mu &= \frac{1}{\sqrt{2}}\left[\frac{\sqrt{1+6w+w^2}}{1+w}\left(\frac{r}{R}\right)^{-\frac{2w}{1+w}}(\partial_t)^\mu \pm \frac{1+w}{\sqrt{1+6w+w^2}}(\partial_r)^\mu  \right],
\end{aligned} 
\ee
normalized such that $k_+ \cdot k_- = -1$.
The expansion parameters of the congruences are
\be 
\theta_\pm[\chi] = \pm \frac{\sqrt{2}(1+2w)}{r\sqrt{1+6w+w^2}},\label{eq:thetas}
\ee
and the intrinsic Ricci scalar of $\chi$ is ${\cal R}[\chi] = 2/r^2$.

Following Ref.~\cite{Nomura:2018aus}, we define the parameter
\be
\varrho = -\frac{{\cal R}[\chi]}{\theta_+[\chi]\theta_-[\chi]}.
\ee
Then the outer entropy $S^{(\text{outer})}[\chi]$ is nonzero when $\varrho > 1$; upon substituting in \Eq{eq:thetas}, this requires $0<w<2/3$.
When this condition is satisfied, the outer entropy is
\be
S^{(\text{outer})}[\chi] = \frac{A[\chi]}{4}\left(1-\frac{1}{\varrho}\right)^2 = \frac{\pi (2-3w)^2 w^2 r^2}{(1+6w+w^2)^2} .
\ee

As shown in Ref.~\cite{Bousso:2018fou}, $S^{(\text{outer})}[\chi]$ can be used to define a quasilocal mass $M^{(\text{outer})}[\chi]$ by setting the outer entropy equal to the Bekenstein-Hawking entropy of a black hole of mass $M^{(\text{outer})}[\chi]$ (see related discussion of the Bartnik-Bray inner mass in Refs.~\cite{Wang:2020vxc,bray2001}). That is, $M^{(\text{outer})}[\chi]$ is the mass of the largest black hole that can be fit inside $\chi$ while keeping the external geometry fixed. This quasilocal mass possesses many nice properties, including positivity, monotonicity under inclusion, conservation, binding energy, and reproducing the irreducible mass when $\chi$ is marginally trapped.
For our chosen $\chi$, the quasilocal mass in this spacetime is
\be
M^{(\text{outer})}[\chi] = \frac{(2-3w)w r}{2(1+6w+w^2)} = \frac{1}{4}(2-3w) m(r),
\ee
for $w\in (0,2/3)$ and where we defined $m(r)$ in \Eq{eq:solution}.
As we increase $r$, $M^{(\text{outer})}[\chi]$ grows as required, and as in the example of the pressureless dust in AdS considered in Ref.~\cite{Bousso:2018fou}, we have $M^{(\text{outer})}[\chi]<m(r)$.

\subsection{Double copy}\label{sec:doublecopy}
The Einstein-scalar action in \Eq{eq:Einsteinscalar} (along with the Kalb-Ramond two-form) appears in the gravity/Yang-Mills double copy~\cite{Bern:2019prr}.
Indeed, on shell, the Ricci scalar factorizes, $R_{\mu\nu} = \partial_\mu \Phi \partial_\nu \Phi$, so the solution in Eqs.~\eqref{eq:Einsteinscalareoms}, \eqref{eq:metscalar}, and \eqref{eq:phisol} can be written in a form that exhibits twofold factorization of indices~\cite{Cheung:2016say} into left and right copies.
Specifically, choosing a field basis in which $\sqrt{-g}g^{\mu\nu} = (e^{-h})^{\mu\nu}$, the left and right indices of $h_{\mu\nu}$ can be labeled as barred and unbarred.
Then as shown in Ref.~\cite{Cheung:2016say}, the Einstein-dilaton action in \Eq{eq:Einsteinscalar} can be written in manifestly double copy-compliant form, from which we can derive equations of motion on which left and right indices manifestly factorize.
However, doing so would require going to a fixed-modulus gauge, where $\partial_\mu \sqrt{-g} = 0$, which means that the coordinates along the angular directions will not be the usual spherical coordinates as in \Eq{eq:metscalar}.

\subsection{Swampland distance conjecture}\label{sec:distance}
In the $w=1$ case, where a source for the energy-momentum tensor is given by the dilaton solution in Eqs.~\eqref{eq:Einsteinscalareoms}, \eqref{eq:metscalar}, and \eqref{eq:phisol},
the form of $\phi \propto t$ suggests a connection between the naked singularity found in \App{sec:singularity} and the swampland distance conjecture~\cite{Ooguri:2006in}.
Consider the radial geodesic solution in \Eq{eq:radw1}.
An observer along this trajectory moving from $r=R$ to $r=0$ would find that $\Phi$ satisfies (in $G=1$ units):
\be 
\Phi = \frac{1}{2\sqrt{\pi}} {\rm arcsech}(r/R),
\ee
which diverges to infinity as $r\rightarrow 0$.
This behavior is reminiscent of the distance conjecture violation seen in bubble-of-nothing solutions~\cite{Draper:2019utz}.
Under the distance conjecture, allowing $\Phi$ to traverse an infinite distance in moduli space would mean that an infinite tower of states becomes massless, taking us out of the effective field theory of \Eq{eq:Einsteinscalar}.
Suppose we merely want to access a region of high curvature, say with $R_{\mu\nu\rho\sigma}R^{\mu\nu\rho\sigma} = \Lambda^4$ for some mass scale $\Lambda$. For $\Lambda R \gg 1$, by \Eq{eq:Kretschmann} we have
\be
\Phi \simeq \frac{1}{2\sqrt{\pi}} \left(\log 2 - \frac{1}{4}\log 3 + \log \Lambda R  \right).
\ee
Thus, under the refined distance conjecture~\cite{Klaewer:2016kiy}, given a tower with some initial mass $M_{\rm i}$, once we probe curvature scales of order $\Lambda$, the tower instead has a (much lighter) mass of $M_{\rm f} \sim M_{\rm i}/\Lambda R$, up to $O(1)$ coefficients. 
In order to trust our solution to scale $\Lambda$, we require that $M_{\rm f} \gg \Lambda$, which means that $\Lambda \ll \sqrt{M_{\rm i}/R}$.
One typically imagines $M_{\rm i}$ to be of order the string scale, so in this geometry the distance conjecture prevents the probing of quantum gravitational degrees of freedom using the high-curvature region, while remaining within the effective theory.
The distance conjecture thus obscures the naked singularity in the fluid solution, in the $w=1$ construction with the dilaton. 
This connection between a swampland condition and a singularity is reminiscent of how the weak gravity conjecture rescues cosmic censorship in the solutions of Ref.~\cite{Horowitz:2016ezu}.
However, in order for our dilaton solution to represent a true violation of the cosmic censorship conjecture, one would require  formation of the geometry in \Eq{eq:metscalar} from generic, smooth initial data~\cite{Wald}, to which the tuning present in Choptuik critical collapse~\cite{Choptuik} could present an obstacle.

\vspace{-1mm}
\section{Conclusions}\label{sec:conclusions}
In this work, we have examined a set of remarkable singular solutions to the TOV equations---the SIS---with metric given in \Eq{eq:metric}, describing a general relativistic perfect fluid with fixed equation-of-state parameter $w=p/\rho$.
Examples of such solutions can be constructed in field and string theories, e.g., a massless scalar with $w=1$.
These solutions possess interesting features in common with black holes, notably, a fixed compactness $M/R$ independent of radius (\Sec{sec:solution}), ability to trap certain geodesics (\App{sec:geodesics}), and a singularity at $r=0$ (\App{sec:singularity}).
However, these objects notably deviate from black holes in crucial ways; in particular, they possess no horizons and thus violate cosmic censorship (\App{sec:singularity}).
We investigated perturbations of these geometries and calculated their tidal Love numbers (\Sec{sec:perturbations}).
Notable aspects of these solutions include the striking behavior of null geodesics, which form photon spheres and the logarithmic spirals explored in \App{sec:geodesics}.
For small $w$, these perfect fluid objects also suggest a general relativistic treatment of dark matter halos, with the $1/r^2$ form of the density profile leading to a flat rotation curve (\Sec{sec:rotationcurve}), though the problem of the transition from fluid to vacuum outside the SIS would need to be solved in a smooth way in order for this to represent a physically-viable solution.

This paper leaves multiple compelling avenues for future work.
Formation mechanisms for these solutions bear further examination.
The naked singularity, leading to geodesic incompleteness and arbitrarily high curvatures that can be seen from asymptotic infinity, raises the question of cosmic censorship and suggests an intriguing connection to the swampland program in string theory (\Sec{sec:distance}); making this link robust would require finding evolution from generic smooth initial data to the solution in \Eq{eq:metscalar}.
Pursuing applications of these metrics to astrophysical models of dark matter and generalizing them to accommodate more complicated dark matter profiles would also be useful, in particular, finding ways of smoothing the transition from the $1/r^2$ fluid to the vacuum region.
Finally, the connection with the double copy discussed in \Sec{sec:doublecopy}, in particular finding an explicitly index-factorized solution of this form, is another potential direction for future work.
Over a hundred years after the discovery of general relativity, the investigation of new solutions to the Einstein equations remains an important avenue for progress, leading to surprising connections among topics in particle physics, astrophysics, and mathematics.
\pagebreak

\begin{center} 
{\bf Acknowledgments}
\end{center}
\noindent I thank Cliff Cheung and Gary Horowitz for discussions, as well as the referees for their useful comments.  
This work was supported at the Kavli Institute for Theoretical Physics by the Simons Foundation (Grant No.~216179) and the National Science Foundation (Grant No.~NSF~PHY-1748958) and at the University of California, Santa Barbara by the Fundamental Physics Fellowship.

\appendix

\section{Singularity}\label{sec:singularity}

\begin{figure}[t]
\begin{center}
\includegraphics[width=10cm]{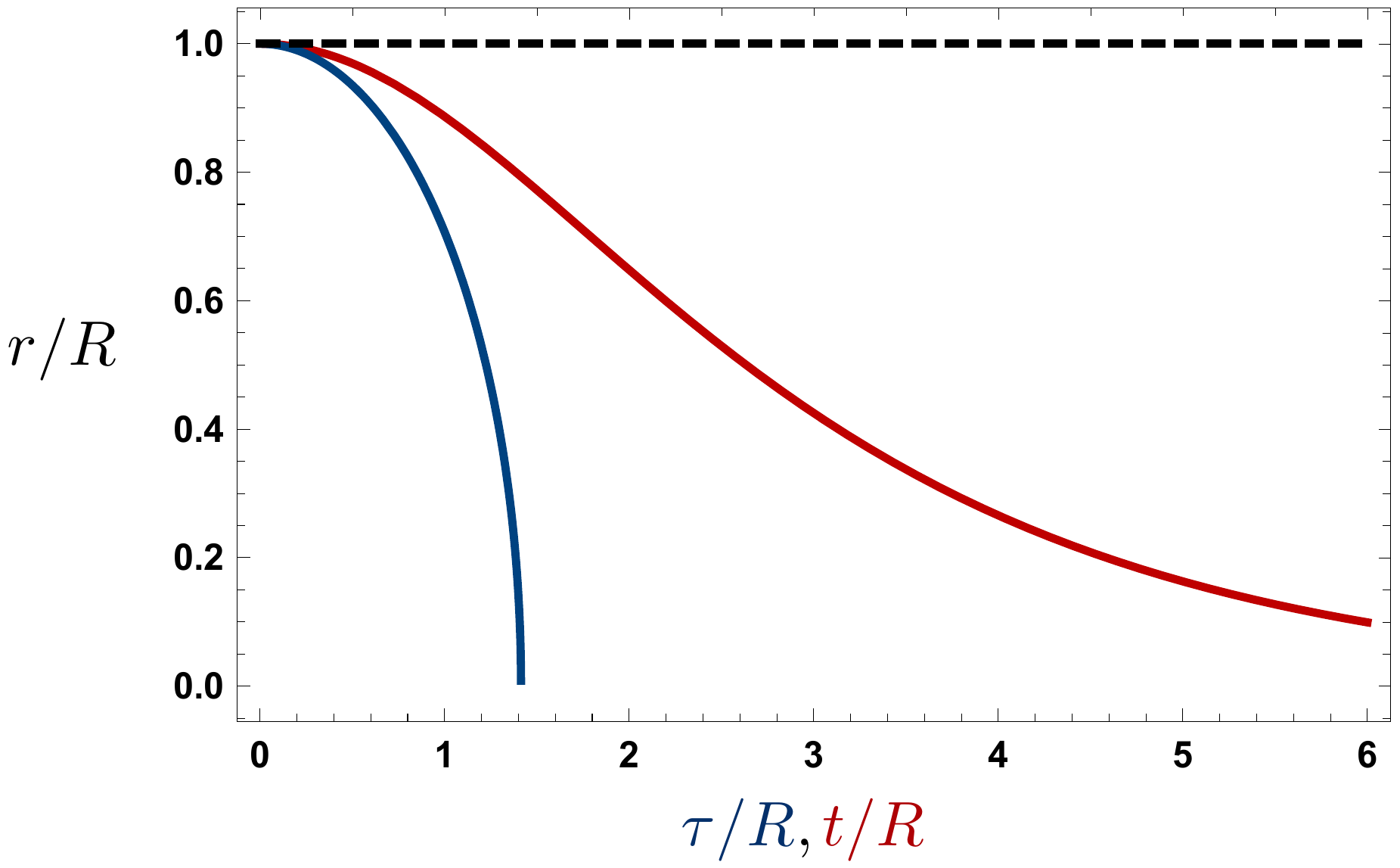}
\end{center}\vspace{-7mm}
\caption{Radial geodesic starting from rest at $r=R$, plotted in terms of proper time $\tau$ (blue) and coordinate time $t$ (red) for the equation-of-state parameter $w=1$.
}
\label{fig:radial}
\end{figure}

For these perfect fluid spacetimes, the density and pressure both diverge as $r\rightarrow 0$, while $g_{tt} \rightarrow 0$ in \Eq{eq:metric}. This suggests the presence of a singularity. First, let us verify that removing the point at $r=0$ results in geodesic incompleteness. 
The energy equation \eqref{eq:EE} derived in \App{sec:geodesics} implies that $\dot t \propto (r/R)^{-\frac{4w}{1+w}}$,  so to an observer outside at $r>R$, measuring time in $t$ coordinates, the clock of the infalling observer will appear to slow down and stop as $r\rightarrow 0$, redshifting as for an observer approaching the horizon of a black hole.
For a geodesic with $r(0)=R$ and $\dot r (0)=v$, the solution satisfies
\be
|\tau \! - \! \tau_0| = \frac{1+w}{1+3w}\frac{R y(\tau)}{\sqrt{1-2C+v^2}}\;\left| {}_2 F_1\!\left[\frac{1}{2},\frac{1}{4}\left(\! 3\! +\! \frac{1}{w}\right);\frac{1}{4}\left(\! 7\! +\! \frac{1}{w}\right);\frac{1-2C}{1-2C+v^2}[y(\tau)]^{\frac{4w}{1+3w}}\right]\right|,
\ee
where
\be
\tau_0 = -R\frac{1+w}{1+3w}\frac{{\rm sgn}\,v}{\sqrt{1-2C+v^2}}\;{}_2 F_1\!\left[\frac{1}{2},\frac{1}{4}\left(3+\frac{1}{w}\right);\frac{1}{4}\left(7+\frac{1}{w}\right);\frac{1-2C}{1-2C+v^2}\right]
\ee
is the proper time at which the geodesic reaches $r=0$. Positive real solutions for $\tau_0$ exist for various choices of $v$, for example, by taking $v=-\sqrt{1-2C}<1$.
If we set $w=1$ and choose $v=0$, we have 
\be 
\begin{aligned}
t(\tau)&= 2R\, {\rm arctanh}(\tau/\sqrt{2}R)\\
r(\tau) &= R\sqrt{1-\frac{\tau^2}{2R^2}},
\end{aligned}\label{eq:radw1}
\ee
and $r=0$ is reached at $\tau = \sqrt{2}R$ (though this takes an infinite time in $t$ coordinates); see \Fig{fig:radial}.
Thus, the spacetime is geodesically incomplete.
Accordingly, there is a naked singularity at $r=0$, since signals from that point can reach the surface at $r=R$ in finite proper time and hence can be sent to infinity without obstruction.
Indeed, singularities without horizons have been shown to be a generic feature of TOV solutions found by imposing conditions at a finite radius and integrating to the center~\cite{Anastopoulos:2020mrt}; in dynamical, astrophysical contexts, well known naked singularities also arise in TOV solutions for collapse~\cite{Christodoulou,Joshi:1993zg,OriPiran}.

We now wish to diagnose the type of singularity present at $r=0$. Inside the star, the Ricci scalar is $4w(1-3w)/[r^2 (1+6w+w^2)]$, which also diverges unless $w=1/3$, corresponding to a photon gas.
A better diagnostic of a curvature singularity is the Kretschmann scalar, 
\be
R_{\mu\nu\rho\sigma}R^{\mu\nu\rho\sigma} =  \frac{16w^2(7+2w+3w^2)}{r^4(1+6w+w^2)^2},\label{eq:Kretschmann}
\ee
which diverges at $r=0$ for any $w>0$, albeit not as strongly as the Schwarzschild singularity, for which $R_{\mu\nu\rho\sigma}R^{\mu\nu\rho\sigma}\propto 1/r^6$.
In the classification of Ref.~\cite{Wald}, $r=0$ corresponds to a type (i) (i.e., scalar curvature) singularity.

The region over which this curvature becomes large actually  has smaller spacetime volume than the flat-space expectation. Taking a spatial ball $B$ of proper area $4\pi r_0^2$ for $r_0<R$ centered at the origin times an interval $T$, $t\in [-t_0/2,t_0/2]$, let us compare the proper spacetime volume ${\cal V} = \int_{B\times T} \sqrt{-\det g}\, {\rm d}t\,{\rm d}r\, {\rm d}\Omega$ to its flat-space analogue ${\cal V}_0 = 4\pi r_0^3 t_0/3$. We find a volume deficit:
\be
\frac{\cal V}{{\cal V}_0} = \frac{1+w}{1+\frac{5}{3}w}\left(\frac{r_0}{R}\right)^{\frac{2w}{1+w}}.\label{eq:volume}
\ee
Nonetheless, this volume is not zero for any $r_0 \neq 0$.  In contrast, purely spatial volumes are larger than in flat space: $|B|/(4\pi r_0^3/3) = 1/\sqrt{1-2C}$. For any chosen large value of the curvature, there is a finite volume of spacetime over which the curvature exceeds that value.

Like the singularity in Reissner-Nordstr\"om spacetime, the perfect fluid solution's singularity is timelike for $0<w<1$, that is, it is a spatial location. This is implied by the fact that timelike geodesics can reach it in finite proper time, and it accords with our intuition since it is located at $r=0$ and the metric does not flip signature. While we observe that setting $r=0$ and ${\rm d}r=0$ in \Eq{eq:metric} results in a vanishing line element, as in the case of a null singularity, this is in fact due to the shrinking of the volume element seen in \Eq{eq:volume}, rather than a null signature for the singularity itself.
To clarify the situation, we can define Kruskal-like coordinates:
\be 
\begin{aligned}
T &= f(r) \sinh(t/R) \\
X &= f(r) \cosh(t/R)
\end{aligned}\label{eq:Kruskaltransform}
\ee
where
\be
f(r) =  \exp\left[\frac{1+6w+w^2}{(1-w)(1+w)}\left(\frac{r}{R}\right)^{\frac{1-w}{1+w}} \right],\label{eq:ff}
\ee
in terms of which the metric for $r\leq R$ can be written as
\be
{\rm d}s^2 = \frac{(1+w)^2}{1+6w+w^2} \left(\frac{r}{R}\right)^{\frac{4w}{1+w}} \frac{R^2}{f(r)^2} \left(-{\rm d}T^2 + {\rm d}X^2\right) + r^2 {\rm d}\Omega^2.
\ee
Radial null geodesics satisfy ${\rm d}X/{\rm d}T = \pm 1$. The singularity at $r=0$ corresponds to $T^2 - X^2 = -1$ and is therefore timelike as expected, for $0<w<1$.

The coordinate transformation function \eqref{eq:ff} is singular at $w=1$, so we must treat that case separately.
There, we can define $f(r) = r^2/R^2$, so that with \Eq{eq:Kruskaltransform} we find
\be
{\rm d}s^2 =  \frac{R^4}{2r^2}(-{\rm d}T^2 + {\rm d}X^2) + r^2 {\rm d}\Omega^2.
\ee
In these coordinates, $T^2 - X^2 = -(r/R)^4$, so the singularity at $r=0$ is null for $w=1$.
This accords with a feature of the $w=1$ geometry in \Eq{eq:metscalar}: at fixed angle, the metric describes two-dimensional Rindler space.

We also note that the metric in \Eq{eq:metric} is conformally flat on spacelike slices of fixed $t$. To see this, let us define an isotropic coordinate $z = (r/R)^\alpha$, where $\alpha = \sqrt{1+6w+w^2}/(1+w)=1/\sqrt{1-2C}$. Then for $r\leq R$, the metric becomes
\be
{\rm d}s^2 =  -(1-2C)z^{2(\sqrt{1-2C}-\sqrt{1-4C})} {\rm d}t^2 + R^2 z^{2(\sqrt{1-2C}-1)} ({\rm d}z^2 + z^2 {\rm d}\Omega^2).
\ee
Similarly, other isotropic coordinates can be defined for the Schwarzschild part of the spacetime,  $r>R$~\cite{Wald}.

\section{Geodesics}\label{sec:geodesics}

To gain a better understanding of these objects, let us investigate the structure of orbits in these spacetimes, focusing on the interior region of \Eq{eq:metric}.
The geodesic equation
$\ddot x^\mu + \Gamma^\mu_{\alpha\beta} \dot x^\alpha \dot x^\beta = 0$
implies
\be 
\begin{aligned}
\ddot t + \frac{4w}{r(1+w)} \dot t \dot r &= 0\\
\ddot r + \frac{2w(1+w)^3}{r(1+6w+w^2)^2} \left( \frac{r}{R}\right)^{\frac{4w}{1+w}} \dot t^2 -\frac{r(1+w)^2}{1+6w+w^2} (\dot\theta^2 + \dot\phi^2 \sin^2 \theta)&=0 \\
\ddot \theta + \frac{2}{r}\dot r \dot\theta - \dot\phi^2  \sin\theta\cos\theta &=0 \\
\ddot \phi + \frac{2}{r} \dot r \dot\phi  + 2 \dot\theta\dot\phi \cot \theta &=0,
\end{aligned}\label{eq:geos}
\ee
where $\dot{}$ denotes differentiation with respect to affine parameter $\lambda$.
By spherical symmetry, we can fix $\theta=\pi/2$ without loss of generality.
The fourth equation in \Eq{eq:geos} is simply conservation of angular momentum,
\be 
L = r^2 \dot\phi = \text{constant},\label{eq:LL}
\ee
and the first equation implies conservation of energy,
\be 
E = \frac{(1+w)^2}{1+6w+w^2}\left(\frac{r}{R}\right)^{\frac{4w}{1+w}} \dot t = \text{constant}.\label{eq:EE}
\ee
These two constants of motion correspond to Killing vectors of the geometry: $(\partial_\phi)^\mu \dot x_\mu = L \sin ^2 \theta$ and $(\partial_t)^\mu \dot x_\mu = -E$.

\begin{figure}[t]
\begin{center}
\includegraphics[width=11cm]{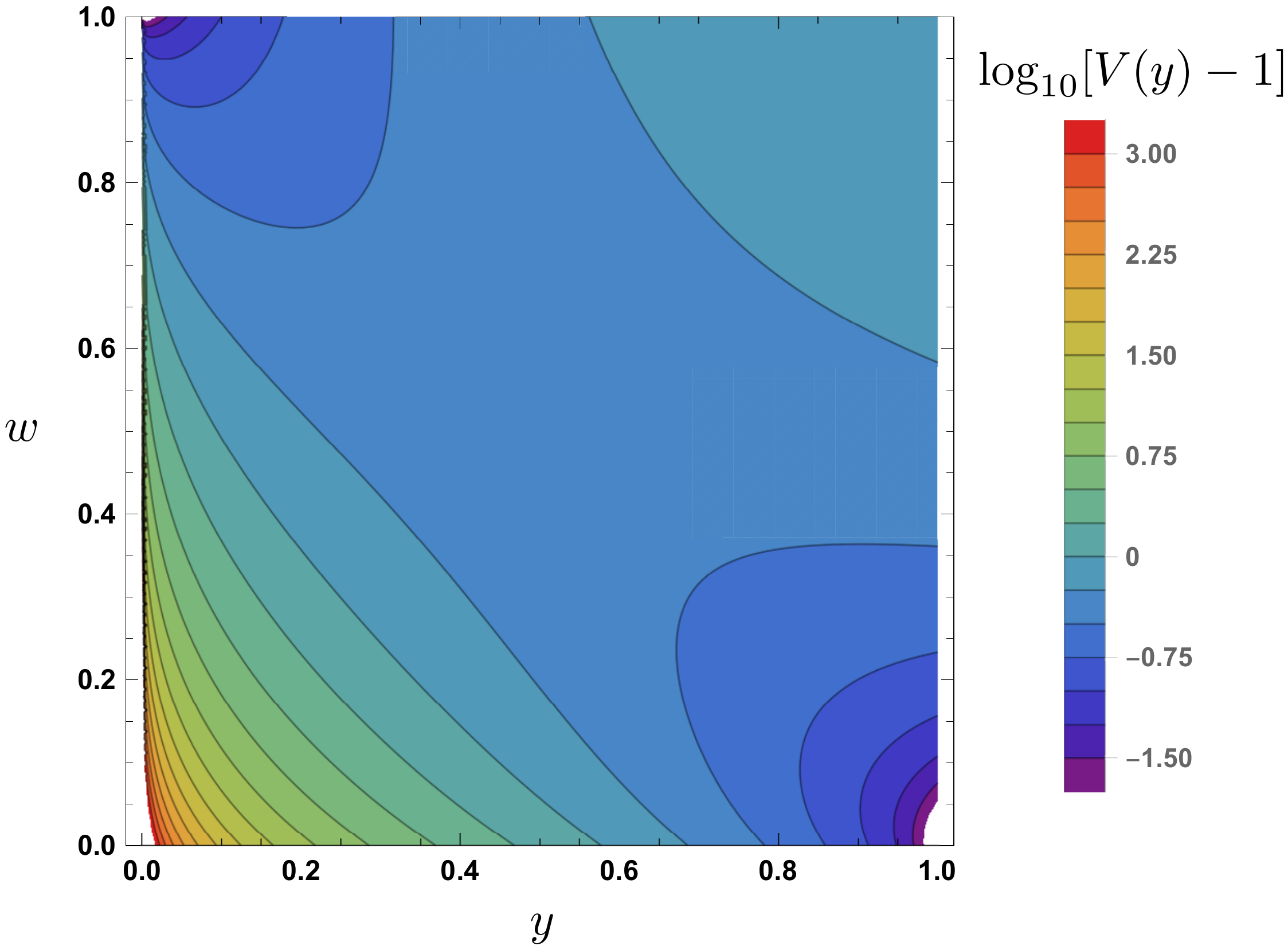}
\end{center}\vspace{-7mm}
\caption{Effective potential $V(y)$ from \Eq{eq:V} for timelike orbits in the singular fluid geometry of \Eq{eq:metric}, in terms of radial coordinate $y$ defined in \Eq{eq:coordinate}. The potential is plotted in units where $R=1$ and for the choice $L=1$. For $w=0$, the angular momentum barrier takes its standard form, but softens as $w$ increases and disappears when $w=1$.
}
\label{fig:potential}
\end{figure}

Let us define the signature of our geodesic:
\be
\epsilon = g_{\mu\nu} \dot x^\mu \dot x^\nu = \text{constant}, \label{eq:signature}
\ee
where $\epsilon = -1,0,+1$ for timelike, null, or spacelike orbits.
Rewriting \Eq{eq:signature} using Eqs.~\eqref{eq:LL} and \eqref{eq:EE}, we have
\be
\dot r^2 +\frac{(1+w)^2}{1+6w+w^2}\left(-\epsilon+\frac{L^2}{r^2}\right) =  \left(\frac{r}{R}\right)^{-\frac{4w}{1+w}} E^2,\label{eq:rad}
\ee
which upon differentiation yields the second geodesic equation in \Eq{eq:geos}. Defining a new radial coordinate, 
\be
y = \left( \frac{r}{R}\right)^{\frac{1+3w}{1+w}},\label{eq:coordinate}
\ee
our radial equation \eqref{eq:rad} becomes
\be
\frac{1}{2}\dot y^2 + V(y) =  \frac{E^2}{2R^2}\left(\frac{1+3w}{1+w}\right)^2,
\ee
where the effective potential is
\be
V(y) = \frac{1}{2R^2}\frac{(1+3w)^2}{1+6w+w^2}\left[-\epsilon y^{\frac{4w}{1+3w}} +\frac{L^2}{R^2 y^{\frac{2(1-w)}{1+3w}}}\right] .\label{eq:V}
\ee
A few comments on the potential are in order. First, when $w\rightarrow 0$, $y\rightarrow r/R$ and we recover the Newtonian limit as expected,
\be
V(y)|_{w=0} =  \frac{1}{2R^2}\left(-\epsilon + \frac{L^2}{R^2 y^2}\right).
\ee
However, whenever $0<w< 1$, the angular momentum barrier in the potential \eqref{eq:V} is softer, i.e., less peaked, than its counterpart in Newtonian mechanics or in the Schwarzschild metric.
When $w=1$, in fact, the barrier disappears completely, 
\be
V(y)|_{w=1} =   \frac{1}{R^2}\left(-\epsilon y +\frac{L^2}{R^2}\right),
\ee
and taking $w>1$ would make the centrifugal potential attractive. See \Fig{fig:potential} for an illustration.

\begin{figure}[t]
\begin{center}
\includegraphics[width=\textwidth]{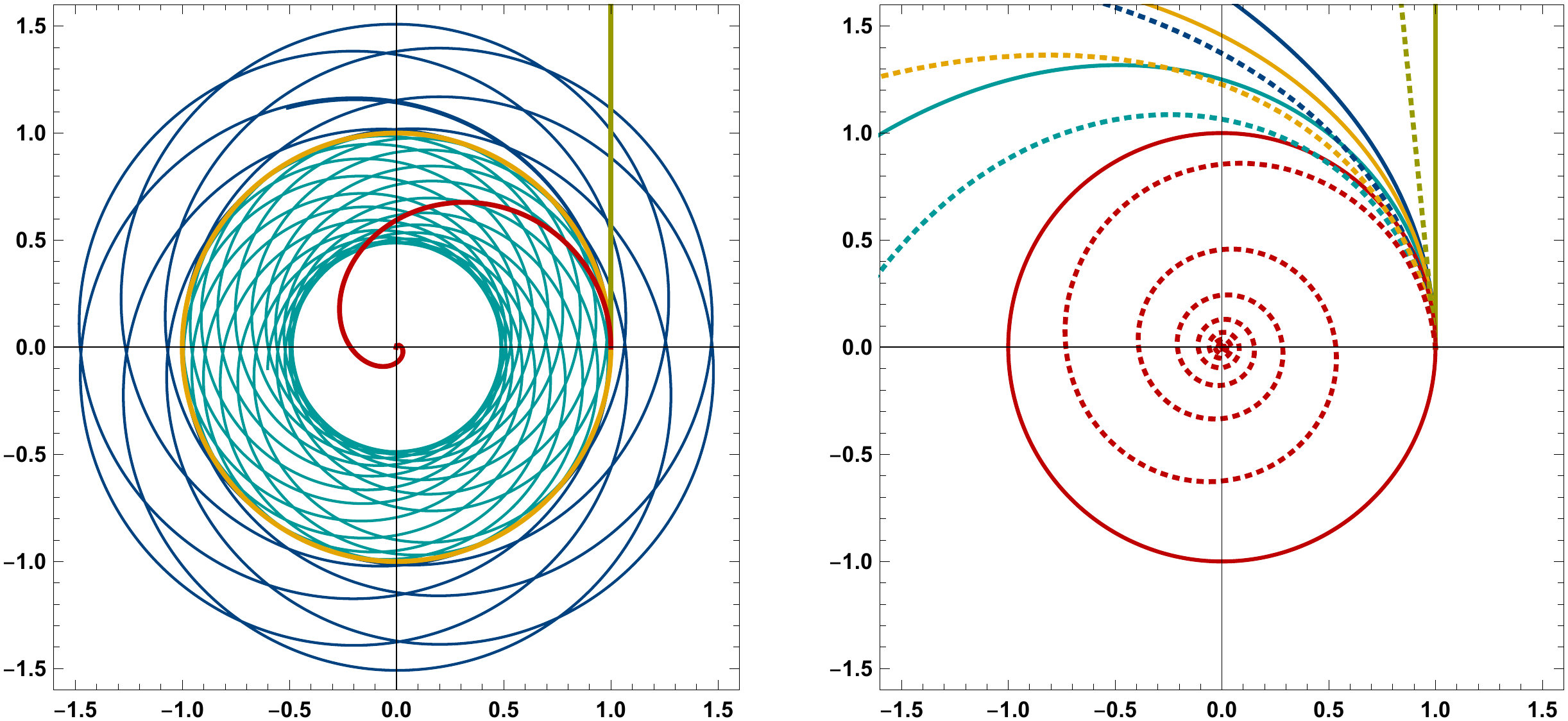}
\end{center}\vspace{-7mm}
\caption{Timelike (left) and null (right) geodesics described by \Eq{eq:geos} for various choices of equation of state: $w=0$ (green), $w=1/4$ (blue), $w=1/3$ (orange), $w=1/2$ (teal), and $w=1$ (red).
Orbits are plotted in the $r$ coordinate in units where $R=1$, with initial radius $r=1$, angular momentum $L=1$, and initial $\dot r(\lambda) = 0$ (timelike and solid lines for null plot) or $\dot r(\lambda) = -0.1$ (dotted lines for null plot). 
For this plot, we allow the $r\leq R$ portion of the geometry in \Eq{eq:metric} to also extend to $r>R$.
}
\label{fig:geodesics}
\end{figure}

The non-$y^{-2}$ form of the centrifugal potential for general $w$ means that orbits will not be elliptical, but instead will precess.
For a timelike circular orbit, the equilibrium condition is
\be 
y_0^{\frac{2(1+w)}{1+3w}} = \frac{1-w}{w}\frac{L^2}{2R^2}.
\ee
Taking $y(\tau) = y_0 + \delta y(\tau)$, we have at leading order in $\delta y$:
\be 
\delta y''(\tau) = -\frac{8w^2(1+w)}{L^2(1-w)(1+6w+w^2)}\delta y(\tau),
\ee
giving an angular frequency of oscillation in $\delta y$ of 
\be
 \Omega_{\rm precession} = \Omega_0\sqrt{\frac{2(1+w)(1-w)}{(1+6w+w^2)}},
\ee
where the orbital frequency is
\be
\Omega_0 =  \dot\phi = \left.\frac{L}{r^2}\right|_{y=y_0}= \frac{2w}{(1-w)L} .
\ee
In the extreme case where $w=1$, stable orbits are not possible: as shown in \Fig{fig:geodesics}, a would-be orbit, with small enough initial velocity that it does not reach the $r=R$ boundary, ends up spiraling down to $r=0$. 

Similarly, in the null case where $\epsilon = 0$, we find that when $w=1$, the orbits of null geodesics form self-similar logarithmic spirals, $\log[r(\lambda)/r(0)] = r(0)\dot r(0) [\phi(\lambda)-\phi(0)]/L$, similar to those found in the collapsing case of Ref.~\cite{OriPiran}. In the $\dot r(0)=0$ case, a photon sphere forms at arbitrary radius; see \Fig{fig:geodesics}.

\bibliographystyle{utphys-modified}
\bibliography{TOV}

\end{document}